\documentclass[twocolumn,pre,nofootinbib,showpacs]{revtex4-1}

\usepackage{amsthm}
\usepackage{epsfig}
\usepackage{amsmath}
\usepackage{graphicx}
\usepackage{color}
\usepackage[table]{xcolor}
\usepackage{dcolumn}% Align table columns on decimal point
\usepackage{bm}% bold math
\usepackage{bbm}

\newcommand{\vek}[1]{\mbox{\boldmath$#1$}}
\newcommand{\change}{}

\usepackage{epsfig}
\usepackage{hyperref}

\begin{document}

\title{Bayesian inference with information content model check for Langevin equations}

\author{Jens Krog}
\author{Michael A. Lomholt}
\affiliation{MEMPHYS - Center for Biomembrane Physics, Department of Physics, Chemistry, and Pharmacy, 
University of Southern Denmark, 5230 Odense M, Denmark}
\date{\today}

\begin{abstract}
The Bayesian data analysis framework has been proven to be a systematic and effective method of parameter inference and model selection for stochastic processes. In this work we introduce an information content model check which may serve as a \textit{goodness-of-fit}, like the $\chi^2$ procedure, to complement conventional Bayesian analysis. We demonstrate this extended Bayesian framework on a system of Langevin equations, where coordinate dependent mobilities and measurement noise hinder the normal mean squared displacement approach.
\end{abstract}


\maketitle
 
\section{Introduction}

Stochastic processes are commonly used for modeling of single particle tracking (SPT) experiments in biophysics. The motion of the particle is assumed to be partly diffusive, but the possibility of effects such as active transport and drift due to flow or some effective potential often cannot be ruled out a priori. Standard tools to distinguish these effects are the estimation of mean square displacement (MSD) or correlations \cite{qian91,meroz15}.
However, if the environment is heterogeneous then the situation easily becomes too complex for estimators such as the MSD to resolve the mechanisms of the underlying dynamics. 

An alternative approach is offered by Bayesian analysis \cite{mackay03,jaynes03,sivia06}, which gives a systematic method for comparing different models of the investigated particle dynamics. The approach simultaneously provides a way of selecting the most probable model among different candidates and provides the most probable associated parameter values together with their uncertainties. A Bayesian approach to analysing SPT trajectories has previously been applied to model selection among hidden Markov models \cite{das09,persson13,monnier15,sgouralis17}, diffusion in varying potentials \cite{masson09,ensign10,turkcan12}, and parameter estimation for optical traps \cite{bera17}, to name just a few examples.

While the Bayesian approach is excellent at choosing the best among specified candidate models and their parameters it does not provide a measure of how well the specified models actually describe the data, i.e., it lacks \textit{goodness-of-fit} measures like the $\chi^2$ procedure of least-squares fitting.

In this article we demonstrate how Bayesian inference can be applied efficiently on SPT data with models that takes the form of a Langevin equation using the nested sampling algorithm introduced by Skilling \cite{skilling04}. The Langevin equation form allows for position or time dependent forces as well as mobilities. Additionally we supplement the Bayesian approach by coupling it with a goodness-of-fit test that checks whether typical trajectories generated from the models using typical inferred parameters reproduce the information content of the observed trajectories at different time scales. The approach is implemented in Matlab, and the code is available at GitHub \cite{sdegit}.

The article is organised as follows. In Section \ref{sec:II} we introduce the general Langevin equation on which we apply the Bayesian inference and information content model check as explained in Section \ref{sec:III}. In Section \ref{sec:IV} we demonstrate how the approach works on synthetic data with a position dependent mobility and we finally conclude in Section \ref{sec:V}.

\section{Generic Stochastic Differential equations}\label{sec:II}
We will assume that a theoretical analysis of the system at hand has resulted in a stochastic differential equation for an observable $x$ on the form of a Langevin equation
\begin{equation}
\label{LE_org}
\frac{dx}{dt} = a(x,t) + b(x,t)\xi(t).
\end{equation}
Both the deterministic term $a(x,t)$ and the coefficient $b(x,t)$ of the stochastic term should be known functions except for a limited number of unknown parameters. The stochastic nature is supplied through the noise function $\xi(t)$. In the following we will assume that $\xi(t)$ is a \textit{white noise} idealization of the physical noise, such that
\begin{equation}
\langle \xi(t) \xi(t') \rangle = \delta(t-t'),
\end{equation}
with an Ito interpretation when $b(x,t)$ is not a constant \cite{gardiner04}.
The characteristics of the process under investigation will then be determined from the other two functions, $a$ and $b$ in Eq. \eqref{LE_org}.
These functions may take on many forms, depending on the system and observable of interest. 

An example of a system that can be brought on such a form is a particle under the influence of a force $f$ and with a position dependent mobility $\mu(x)$. The probability distribution $p(x,t)$ for the particle position $x$ at time $t$ obeys a Fokker-Planck equation
\begin{equation}
\label{fpeq}
\partial_t p(x,t) = - \partial_x[\mu(x)fp(x,t) - k_BT\mu(x)\partial_x p(x,t)], 
\end{equation}
where $k_B T$ is Boltzmann's constant times temperature.
This equation generalises the usual high friction limit of Kramers' equation \cite{kampen97} to the situation with position dependent mobility.
The order of the second partial derivative $\partial_x$ relative to $\mu(x)$ is fixed by the condition that the Boltzmann distribution should be a stationary solution when the force $f$ is conservative. We have not indicated a dependence of position or time in the force $f$, since in later sections we will take it to be constant.
The Fokker-Planck equation can be converted to the form of the Langevin equation \eqref{LE_org} giving {\change for the Ito interpretation of the noise} \cite{gardiner04}
\begin{equation}
\label{LE_FP}
\frac{dx}{dt} = \mu(x)f + k_BT\frac{\partial}{\partial x}\mu(x) + \sqrt{2k_BT\mu(x)}\xi(t),
\end{equation}
where the coordinate dependent $\mu(x)$ appears in the deterministic as well as the stochastic term. We only study the one dimensional case here, but generalizations of this equation to higher dimensions are straightforward.

If the noise is assumed to be negligible, then Eq. \eqref{LE_FP} can be used to obtain deterministic equations of motion for $x$, which can be fitted to obtain the correct form of the deterministic terms. On the other hand, if the mobility {\change and the force are} independent of $x$, then mean-squared-displacement techniques may be employed to unveil characteristics of the stochastic dynamics.
The Bayesian inference approach, however, relies on none of these assumptions, and can be used to infer the functional form of $\mu(x)$, and thus the best underlying model, in any case where a Langevin equation on the form \eqref{LE_FP} can be found.

\section{Bayesian parameter inference and model testing}\label{sec:III}
In the following we will refer to a specific functional form of the underlying Langevin equation \eqref{LE_org} as a \textit{model}, $M$. Labeling the different models via a subscript $i$, model selection through the knowledge of some experimentally observed data $\vek{x}$ then follows the logic of Bayes' theorem
\begin{equation}
\label{bayes_thm}
P(M_i|\vek{x}) = \frac{P(\vek{x}|M_i)P(M_i)}{P(\vek{x})}.
\end{equation}
Here, the left hand side is called the \textit{posterior} probability for the model $M_i$ given the data, while $P(\vek{x}|M_i)$ is the \textit{likelihood} of $M_i$. $P(M_i)$ is the \textit{prior} probability of the model, and may be set equal for all models to achieve impartial judgement. The denominator in \eqref{bayes_thm} is independent of the model, and may be seen as a normalization constant.

To compare the posterior probabilities of a set of models, one must then simply calculate the likelihoods $P(\vek{x}|M_i)$. As each model $M_i$ contains a number of parameters $\vek{\theta}$, this is found by marginalizing over the unknown parameters,
\begin{equation}
\label{evidef}
P(\vek{x}|M_i) = \int P(\vek{x}|\,\vek{\theta},M_i)P(\vek{\theta}|M_i)d\vek{\theta},
\end{equation} 
where $P(\vek{x}|\,\vek{\theta},M_i)$ and $P(\vek{\theta}|M_i)$ are the likelihood function and prior probability, respectively of the parameters for the model $M_i$. Considering each possible set of parameters $\vek{\theta}$ as representing a unique model, $Z_i=P(\vek{x}|M_i)$ is often referred to as the evidence for that class of models, and used to compare the relative merit of these classes.

Assuming that the positions $x_0,x_1,\dots,x_N$ constituting $\vek{x}$ are sampled at closely spaced times $t_0,t_1,\cdots,t_N$ such that the force $f$ and mobility $\mu$ vary very little during a time interval we can integrate Eq. \eqref{LE_FP} over each time interval to obtain the likelihood function
\begin{equation}
\label{likelihood}
P(\vek{x} |\vek{\theta},M_i) = \prod_{j=1}^N \frac{1}{\sqrt{2\pi\sigma_j^2}}\,\text{exp}\left[-\frac{\left(x_j -  \overline{x}_j\right)^2}{2\sigma_j^2}\right].
\end{equation}
with the following means and variances for each new position
\begin{align}
\overline{x}_j &\equiv \langle x(t_j) \rangle \nonumber \\
\label{xmean}
&= x_{j-1} + \left[\mu(x_{j-1}) f + k_BT\partial_x\mu(x_{j-1})\right](t_j - t_{j-1}),\\
\sigma_j^2 &\equiv \langle (x_j - \overline{x}_{j})^2 \rangle \nonumber \\
&=
\label{xvar}
 2k_BT\mu(x_{j-1}) \,(t_j - t_{j-1}).
\end{align}

By defining the priors $P(\vek{\theta}|M_i)$ the integrand of Eq. \eqref{evidef} is known, and a numerical integration may be performed to obtain the evidence $Z_i$ and posterior probability
\begin{equation}
\label{thetapost}
P(\vek{\theta}|\vek{x},M_i)=\frac{P(\vek{x}|\,\vek{\theta},M_i)P(\vek{\theta}|M_i)}{Z_i}
\end{equation}
 for the parameters of each model. The evidences, depending on the given likelihoods $P(\vek{x}|\vek{\theta},M_i)$ and on the user defined priors $P(\vek{\theta}|M_i)$, constitute a degree of plausibility of the model $M_i$. An evidence ratio $Z_1/Z_2 = 10$ may be interpreted, assuming impartial priors, as a $91\%$ chance of $M_1$ being the better model of the two. As long as careful attention is given to the process of prior assignment, this estimate will yield reliable conclusions.
{\change
One thing to be aware of, though, is situations where one model is an expansion of another simpler model by the addition of a parameter $\theta$ that can deviate from the fixed value in the simpler model $\theta_{\rm norm}$. If the prior for $\theta$ is very broad in the sense that very little is known about the true value of $\theta$, then the Bayesian analysis may yield the highest evidence for the model with $\theta = \theta_{\rm norm}$, even though the true value is different from $\theta_{\rm norm}$. This reflects the fact that for very broad priors a small deviation will be evaluated as unimportant within the Bayesian framework, until a sufficient amount of data is accumulated. 
Finding relatively small effects requires relatively large amounts of data in any analysis. The dependence of model selection on the broadness of a prior is, however, absent in frequentistic approaches. The discrepancy between the two approaches is called Lindley's paradox and has been discussed extensively. See for instance \cite{cousins17} for a discussion in the context of high energy physics.
}
%{\color{red}One caveat is that if a parameter is added by letting it vary from its previously fixed value, but the prior is very broad in the sense that very little is known about the supposed value, then the Bayesian analysis may very likely yield lower evidence for the model with  }

\subsection{Measurement noise effects}\label{subsec:noise}

If the subject of investigation is a process where an error is connected to the data collection process itself, it may be necessary to distinguish between this measurement noise and the noise native to the underlying stochastic dynamics.

The presence of measurement noise $\vek{\eta}$ shifts the true values $\vek{x}^\text{true}$, such that we observe
\begin{equation}
x_j = x^\text{true}_j + \eta_j,
\end{equation}
where we assume a zero mean Gaussian measurement noise with $\langle \eta_i\eta_j\rangle = \delta_{i,j}\sigma_{\rm mn}^2$. This is appropriate for situations where the coordinate is determined by averaging an intensity profile at each measurement.

If the inherent noise of the system does not depend on the coordinate, we can adjust the mean $\overline{x}_j$ and variance $\sigma_j^2$ of the true coordinates to incorporate the measurement noise through an iteratively calculated mean $\tilde{x}_j$ and variance $\tilde{\sigma}_j^2$ \cite{rasmussen06}. In Appendix~\ref{app_noise} we derive (alternatively, see for instance \cite{rasmussen06}) that for constant mobility $\mu$ and force $f$
\begin{align}
\label{nmean}
\tilde{{x}}_j &= \overline{x}_j - \frac{\sigma_{\rm mn}^2}{\tilde{\sigma}_{j-1}}(x_{j-1} - \tilde{x}_{j-1}) \\ 
\label{nvar}
\tilde{\sigma}_j^2 &= \sigma_j^{2} + \sigma_{\rm mn}^2\left(2-\frac{\sigma_{\rm mn}^2}{\tilde{\sigma}_{j-1}^2} \right),
\end{align}
with the starting values $\tilde{x}_1 = \overline{x}_1$ and $\tilde{\sigma}_1^2 = \sigma_1^2 + 2\sigma_{\rm mn}^2$. The likelihood is then computed by replacing $\overline{x}_j$ and $\sigma_j$ in Eq. \eqref{likelihood} with $\tilde{{x}}_j$ and $\tilde{\sigma}_j$.

If $\mu$ and $f$ are not independent of position $x$, then the evaluation of the the mean and variance given in Eqs. \eqref{xmean} and \eqref{xvar} is in principle not possible, since $\mu$ and $f$ depend on the true positions. However, if $\mu$ and $f$ varies only slowly on the length scale $\sigma_{\rm mn}$ associated with the measurement noise, then we can, to a good approximation, use the observed positions when evaluating Eqs. \eqref{xmean} and \eqref{xvar}. In the following we will assume this weak dependence of $\mu$ and $f$ on position and evaluate them in this way.

{\change We remark that we model the noise as independent between measurements of the position. However, there is a number of effects that can lead to correlations in the measurement noise. For instance, a cameras finite exposure time gives rise to scalded motion blur \cite{berglund10}. Including such effects in the modelling would of course be beneficial if they are expected to be substantial.}

\subsection{Information and model testing}\label{subsec:inf}
So far we have only discussed how the Bayesian setup generates relative degrees of plausibility of different models, but not touched upon the problem of determining whether or not the set of models available actually can match the data well.
A way to test the latter is to compare properties of the data with replicated data from the models. The nested sampling framework offers a readily accessible property: the \textit{Shannon information} \cite{shannon48}.
For a model $M$ we can quantify the information that some observation $\vek{x}$ gives us about the parameters of the model as
\begin{equation}
\mathcal{H} = \int P(\vek{\theta}|\vek{x},M) \ln \frac{P(\vek{\theta}|\vek{x},M)}{P(\vek{\theta}|M)}d\vek{\theta}. 
\end{equation} 
Using 
Eq. \eqref{thetapost}
the corresponding expression takes the form \cite{jaynes67}:
\begin{equation}
\label{information}
\mathcal{H} =\frac{1}{Z(M)} \int P(\vek{x}|\vek{\theta},M)P(\vek{\theta}|M) \text{ln} \left(\frac{P(\vek{x}|\vek{\theta},M)}{Z(M)} \right) d\vek{\theta} .
\end{equation}
This can also be read as the posterior average of minus the information content in the observation, with the factor $Z(M)$ included to make the argument of the logarithm unitless. 

In order to evaluate the merit of a model, we use the fact that the information content should be similar to typical values for replicated observations generated from the model with parameters taken from the posterior. 
We thus have a measure to be used for comparison. By drawing a sample of parameters $\vek{\theta}^*$ from the posterior distribution
\begin{equation}
\label{postdist}
P(\vek{\theta}^*|M,\vek{x}) =  \frac{P(\vek{\theta}^*,\vek{x}|M)}{P(\vek{x}|M)} = \frac{P(\vek{x}|\vek{\theta}^*,M)}{P(\vek{x}|M)}P(\vek{\theta}^*|M).
\end{equation}
and using these to generate a replicated trajectory $\vek{x}^*$ we can calculate
\begin{equation}
\label{inf_measure}
h(\vek{x}^*,\theta^*) = \text{ln} \left(\frac{P(\vek{x}^*|\vek{\theta}^*,M)}{Z(M)}\right),
\end{equation}
where $Z(M)=P(\vek{x}|M)$.
Generating several pairs of $\vek{\theta}^*$ and $\vek{x}^*$ we obtain a cluster of $h$-values. If the models describe the original data $\vek{x}$ well with respect to information content, then $\mathcal{H}$ should be somewhere within this cluster of $h$-values for replicated data. We demonstrate this with examples in Section \ref{sec:IV}.

As an alternative way of representing the check of whether the information content of the original data is typical with respect to the posterior of the Bayesian inference we calculate,
by averaging over many draws of $\vek{\theta}^*$ and corresponding $\vek{x}^*$, 
the posterior predictive $p$ value \cite{gelman13}
\begin{equation}
p=\langle\Theta[h(\vek{x}^*,\vek{\theta}^*) - h(\vek{x},\vek{\theta}^*)]\rangle ,
\end{equation}
where $\Theta(x)$ is the Heaviside function. {\change This $p$ value is the probability of replicated data being more extreme with respect to information content than the observation $\vek{x}$.}
If the original data $x$ is well described by the model, then 
the $p$ value should not be extreme, i.e., improbably close to zero or one.
%, then we conclude that the model describes the data well with respect to information content. 
{\change In particular, if the model contains the true underlying model as a special case and the posterior probability for the parameters is narrowly distributed around their true values, then the above $p$ value should be uniformly distributed between zero and one. However, since the posterior probability distribution is obtained based on the same observational data that is used for the calculation of the $p$ value, then the distribution of $p$ values will tend to concentrate more towards a value of one half \cite{gelman13}. In any case we conclude that the model describes the data well with respect to information content if the $p$ value is not extremely close to zero or one. The information measure regarded here is convenient because it is readily available from the calculations used for the Bayesian analysis. For a more involved procedure addressing a uniformly distributed measure see \cite{hongli2005}.}
Note that this analysis does not require the comparison between different models, but yields a check of the absolute quality of a model in contrast to the relative one yielded by the evidence ratios. If an observed data set is not typical for the model with respect to information content, then the model is likely to miss a key aspect of the underlying process, even though this model has yielded the highest evidence of the ones tested. 

Since the information content of the data can be quite closely related to the evidence of the model, then the model check as described above might not be a very stringent test. {\change We suspect that this is the reason why the authors have not been able to find this particular model check described elsewhere.} To improve on {\change the stringency of the model check}, we additionally rescale the data by only considering every $n$'th data point in the series and in turn generate replicated tracks with correspondingly longer but fewer time steps.
The information content of a reduced track $\vek{x}^*_n$ for parameters $\vek{\theta}^*$ is then calculated as
\begin{equation}
h(\vek{x}^*_n,\theta^*) = \text{ln} \left(\frac{P(\vek{x}^*_n|\vek{\theta}^*,M)}{Z_n(M)}\right),
\end{equation}
where $Z_n(M)=\int P(\vek{x}_n|\vek{\theta},M)P(\vek{\theta}|M)d\vek{\theta}$ with $\vek{x}_n$ being the reduced original data. The $p$ value is then found as before, and the posterior average becomes
\begin{equation}
\label{information_n}
\mathcal{H}_n =\frac{1}{Z(M)} \int P(\vek{x}|\vek{\theta},M)P(\vek{\theta}|M) \text{ln} \left(\frac{P(\vek{x}_n|\vek{\theta},M)}{Z_n(M)} \right) d\vek{\theta} .
\end{equation}
Obtaining these quantities for a number of different $n$ provides a more stringent test that models will fail if their probability distribution do not scale in a proper way with step size compared with the distribution ingrained in the experimentally observed steps. 
{\change We demonstrate this later, for instance for the case of overlooked measurement noise displayed in Fig.~\ref{inf_nonoise}}.
 To make the above reasoning more precise we have detailed it for a case of diffusive Gaussian models in Appendix \ref{H_scale}.

We have framed the discussion in this subsection in the language of information theory, since our inspiration for choosing this particular model check has come from the concept of ``typical set'' \cite{mackay03}: in essence the model check tests whether the observed data belongs to the typical set of the inferred model. Another way to phrase this is, that the check examines whether the observed data belongs to the very many possible trajectories of the model that does not have an unlikely high or small value for their probability density, i.e., the model fails the check if the observed trajectory is an extreme case of high or low probability density relative to typical trajectories produced by the model. 

{\change We remark that there is a number of approaches to model selection involving information criteria, which rank models based on their predictive accuracy as measured by information content of the data \cite{gelman13,burnham02}. Sometimes in these approaches the models are optimized on part of the data and then cross-validated on another part. This is different from the approach in this article, where the full data-set is used for model selection and parameter fitting, and then also used for model checking.}

\section{Implementation and examples}\label{sec:IV}
To demonstrate the Bayesian inference method and the information content model check, we will examine artificial data samples generated by the Langevin equation on the form shown in Eq. \eqref{LE_FP}, with the mobility,
\begin{equation}
\mu(x) = D_0|x|^\alpha,
\end{equation}
similar to what has been theoretically investigated in \cite{cherstvy13}.
$D_0$ has dimension of $\text{length}^{1-\alpha}\text{time}^{-1}\text{force}^{-1}$, and the force $f$ as well as the exponent $\alpha$ will be assumed to be either zero or a nonzero constant. By letting the force be constant while the mobility varies we essentially analyse a situation complementary to the setup investigated in \cite{ensign10}, where the force was assumed to be a polynomial in $x$. 
We stress that any set of functions  $\mu(x),f(x)$ could have been chosen, as long as $\mu(x)$ was differentiable.
By choosing values for $D_0$, $\alpha$, $f$, and $\sigma_{\rm mn}^2$ as well as a starting point $x_0$, we use \eqref{nmean} and \eqref{nvar} to generate the steps iteratively. To specify the assumptions of the models we consider we use the subscripts "pull" and "free" to signify whether or not the model assumes a nonzero force $f$, while "mn" and "clean" specifies whether or not the model takes into account measurement noise. {\change Throughout all examples we set $k_B T=1$.}

To calculate the respective evidences $Z_i = P(\vek{x}|M_i)$, as given in Eq. \eqref{evidef}, we specify the use of a Jeffreys prior for $D_0 \in [10^{-4} ; 10^2]$, such that
\begin{equation}
P(D_0|M_i) = \frac{1}{D_0 \text{ ln}\,10^6}\Theta\left(3 - \left|\text{log}_{10} \left(\frac{D_0}{10^{-1}} \right)\right|\right), 
\end{equation}
allocating the same amount of probability within each decade, while assigning the uniform priors 
\begin{align}
P(\alpha|M_i) &= \tfrac{1}{4}\Theta(2-|\alpha|), \\
P(f|M_{\rm pull}) &= \tfrac{1}{2}\Theta(1-|f|), \\
P(\sigma_{\rm mn}^2|M_{mn}) &= \frac{1}{100}\Theta(100-\sigma_{\rm mn}^2).
\end{align}
{\change
We choose to use the Jeffreys prior for parameters where the correct order of magnitude might not be known beforehand, and uniform priors for parameters that may vanish or be negative, or where 
the order of magnitude may be estimated from the experimental background information.}
If the amount of data is not too small then the specific choice of prior should have negligible influence on the inference results {\change for the parameter values as long as the prior is nonzero} in the peak of the likelihood. {\change As discussed earlier, then the specific choice of prior can influence the result for model selection (Lindley's paradox). Thus care must be taken to make the prior reflect the a priori expectations for the parameters in the specific experiment. For a suggestion of a non-informative approach towards this using reference priors see \cite{berger89}.}

The integral in Eq. \eqref{evidef} is evaluated numerically with the nested sampling algorithm \cite{skilling06}, in which stochastic sampling of the parameter space is performed via a Markov chain Monte Carlo (MCMC) technique. 
From an initial uniform distribution of sample points within parameter space, the sample with the lowest likelihood is progressively removed and replaced by a new sample within the remaining volume of parameter space with higher likelihood. The MCMC technique generates this new sample and ensure that it is independent of the remaining samples and as such is once again uniformly distributed.
Through such sampling, inferred parameter values are automatically included through the estimated posterior probability landscape. The inferred posterior distribution can then be used to generate the samples needed for the information content estimates in Eq. \eqref{inf_measure}.

\subsection{Noise free data}

\begin{figure}[t]
\begin{center}
\includegraphics[width = 0.45\textwidth]{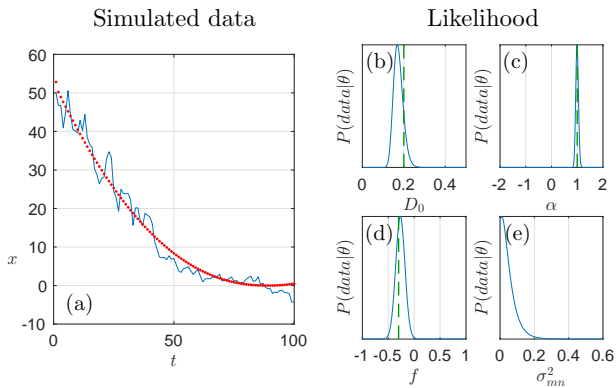}
\caption{(a) Sample track along with a deterministic fit. \, (b)~-~(e) Likelihoods for variations of single parameters (right) around the maximum likelihood point for the best candidate model, which in this case is $M_{\text{pull,clean}}$. The dashed lines represent the true values.}
\label{sampleplot}
\end{center}
\end{figure}

In the left side of Fig.~\ref{sampleplot} an artificial sample track is displayed that we will apply the Bayesian inference method on.
For now we consider data with no measurement noise, so we focus on two models, $M_{\text{free,clean}}$ and $M_{\text{pull,clean}}$, where both $D_0$ and $\alpha$ are free parameters, while $\sigma_{mn}^2 = 0$, but where only the latter model allows a nonzero force. 

Integrating over all of parameter space, we find
\begin{equation}
\text{ln }Z_{\text{pull,clean}} - \text{ln }Z_{\text{free,clean}} = 2.42 \pm 0.15. 
\end{equation}
Thus $Z_{\text{pull,clean}}/Z_{\text{free,clean}} \approx 11.3$, which strongly favors the model with a nonzero force. We obtain uncertainty for the ratio of the evidences via the inferred information \cite{skilling04}. 
For $M_{\text{pull,clean}}$ we obtain the inferred values shown in Table~\ref{inf_param}.
\begin{center}
\begin{table}[h]
\begin{tabular}{|l|r|r|r|}
\hline
Parameter & Inferred value & True value &{\change Fitted value} \\
\hline
$D_0$ & $0.174 \pm 0.039$  & 0.2 & {\change N/A} \\
\hline
$\alpha$ & $1.001 \pm 0.091$ & 1.0 & {\change $0.553 \pm 0.050 $} \\
\hline
$f$ & $-0.277 \pm 0.097$ & -0.3 & {\change N/A} \\
\hline
{\change $D_0 \times f$} & {\change $-0.048 \pm 0.022$} & {\change -0.06} & {\change $-0.148 \pm 0.012$} \\
\hline 
\end{tabular}
\caption{{\change Bayesian-inferred, true and naively fitted} parameters for the data in Fig.~\ref{sampleplot}.}
\label{inf_param}
\end{table}
\end{center}
As expected the parameter values are inferred with reasonable uncertainty.
The plots in the right side of Fig.~\ref{sampleplot} show the development of the likelihood function as the parameters are varied one at a time, showing a slightly skewed distribution for $D_0$ as it is restricted to be positive. This illustrates how Bayesian inference describes the entire parameter space in terms of the associated likelihood of the data. In addition, by examining the likelihood as a function of the noise parameter we see how the best model correctly is the one without noise.

{\change
For comparison, we also estimate the parameters in this example via simple fitting, by removing all (physical) noise terms, i.e. second derivative terms in the Fokker-Planck equation \eqref{fpeq} from which we then recover a deterministic differential equation\footnote{As the noise term is coordinate dependent, the mean squared displacement method of estimating noise cannot be employed.}:
\begin{equation}
\label{detx}
\frac{dx_{det}}{dt} = \mu(x)f = D_0 f|x_{det}|^\alpha,
\end{equation}
where $x_{det}(t)$ is uniquely determined by the initial condition. Solving this equation then leads to an expression for $x_{det}(t)$ which is fitted to the data in Fig.~\ref{sampleplot} via the native \verb+fit+ function in Matlab, showing the best fit in red dots. Note that only the product of $D_0$ and $f$ appears in the deterministic expression \eqref{detx}, such that neither may be fitted individually. As is evident from Table \ref{inf_param}, this naive fitting procedure fails to estimate the parameters correctly although the parameter space has been decreased. This is not surprising, as the simple fitting procedure considers the noise to be independently and identically distributed. In reality, the positions are highly correlated in the fluctuating time series considered here.
As the deterministic approach does not include any knowledge about the fluctuations of the measured quantities, regular $\chi^2$ testing cannot be employed as a model check.
The ability to deal with probabilities for fluctuating data rather than just an averaged curve enables the Bayesian approach to estimate the noise and hence the mobility $\mu$ independently of the force\footnote{The shortcomings for the fitting procedure may of course be remedied by more sophisticated methods, such as a stepwise fitting procedure with changing variance as the one employed in \cite{calderon13},  which, however, requires many more data points. Such techniques take a step towards the Bayesian approach by analysing the maximum likelihood parameters, while not exploring their distribution.}.
}

\begin{figure}[h]
\begin{center}
\includegraphics[width = 0.45\textwidth]{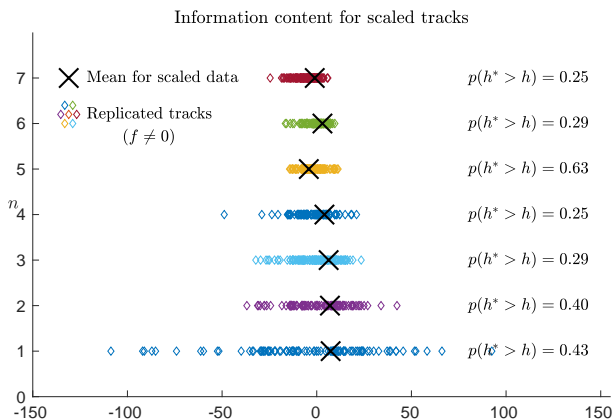}
\caption{The inferred mean information $\mathcal{H}_n$ for the data set in Fig.~\ref{sampleplot} assuming the model $M_{\text{pull,clean}}$ along with the information $h^*=h(\vek{x}^*_n,\vek{\theta}^*)$ for replicated tracks from the same model and corresponding $p$ values.}
\label{inftrue_1}
\end{center}
\end{figure}

Taking the data from Fig.~\ref{sampleplot} and the corresponding inference, we calculate the information for changed timescales by considering only the data points $x_{1 + in}$ for integer $i$. We draw parameters to generate new tracks via Eq. \eqref{postdist} with $M_{\text{pull,clean}}$. The generation of one hundred tracks for each scaling and the associated measures of information $h(\vek{x}_n,\vek{\theta}^*)$ are shown in Fig.~\ref{inftrue_1} along with the mean of the information for the model as given in Eq. \eqref{information_n}. The value of $Z_n(M)$ is estimated with prior samples from the nested sampling run on the non-scaled original data.
Any error in this estimate will only cause a shift in the collective information content estimates, leaving all comparisons and \textit{p} values unaffected. 

Evidently, the information measures for the replicated tracks are distributed around the information of the original data such that we conclude that the original data set is within the ``typical set" for the model with respect to information content.

\begin{figure}[h]
\begin{center}
\includegraphics[width = 0.45\textwidth]{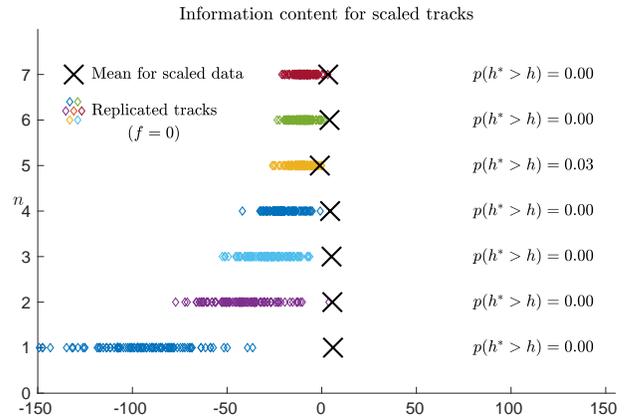}
\caption{Information measures as in Fig. \ref{inftrue_1} for the data set in Fig.~\ref{sampleplot} assuming the model $M_{\text{free}}$.}
\label{inffalse_1}
\end{center}
\end{figure}

On the other hand, if only the model with no force $(M_{\text{free}})$ is used to perform the parameter inference and replicated track generation, the picture in Fig.~\ref{inffalse_1} appears.
When the force is set to zero, the replicated data for the model deviate considerably from the original data in terms of the information. Note that this is the case even though the original analysis has located the best possible available parameters for the model. This warns us that a critical aspect of the process has been overlooked, namely the inclusion of the force.

Since the force parameter adds a drift to the system, the lack of such a parameter might have been easily detectable from visual inspection of the data alone. Conversely, a significant measurement noise might not be detected by simple inspection. If this effect is neglected in the analysis however, the comparison of the information of the original data versus that of replicated data will reveal the flaw in the model.

\subsection{Noisy data}
\begin{figure}[h]
\begin{center}
\includegraphics[width = 0.45\textwidth]{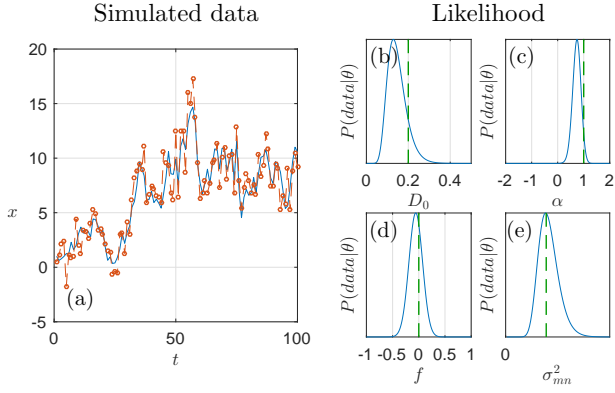}
\caption{(a) A clean sample track in solid blue along with the same track with measurement noise pollution in orange circles. (b)~-~(e) Likelihoods for variations of single parameters around the maximum likelihood point for the best candidate model ($M_{\text{free,mn}}$). The true values are marked by the green dashes. 
{\change Note that the likelihood curves are one dimensional slices of a four dimensional function.}}
\label{noisesampleplot}
\end{center}
\end{figure}

The generation of noisy data is simply done by generating clean data as above and then adding noise with the appropriate variance afterwards.
In Fig. \ref{noisesampleplot} an artificial data set contaminated with random noise is shown alongside slices of the likelihood function around the inferred maximum likelihood parameters.

By reusing the priors from the previous example to compare the calculated evidences for the models with and without measurement noise ($M_\text{free,clean} ,M_\text{free,mn}$), we find
\begin{equation}
\text{ln }Z_{\text{free,mn}} - \text{ln }Z_{\text{free,clean}} = 10.62 \pm 0.14, 
\end{equation}
such that $Z_{\text{free,mn}}/Z_{\text{free,clean}} \sim 4.1\times 10^{4}$, 
firmly establishing superiority of the noise model.
The inferred parameters for the model are shown in Table~\ref{inf_param_noise}. 
\begin{center}
\begin{table}[h]
\begin{tabular}{|l|r|r|}
\hline
Parameter & Inferred value & True value \\
\hline
$D_0$ & $0.21 \pm 0.22$  & 0.2 \\
\hline
$\alpha$ & $0.94 \pm 0.49$ & 1.0 \\
\hline
$\sigma_{\rm mn}^2$ & $2.49 \pm 0.61$ & 2.0 \\
\hline
\end{tabular}
\caption{Inferred and true parameters for the data in Fig.~\ref{noisesampleplot}}
\label{inf_param_noise}
\end{table}
\end{center}
The inferred diffusion coefficient has a large uncertainty which is expected in the presence of large measurement noise. At this level, however, the true value is still well within the error. 
 {\change Note that the likelihood function for $D_0$ is shown at a specific (the maximum likelihood) point for the remaining parameters, which is why it does not correspond at a one-to-one basis to the estimate given in Table~\ref{inf_param_noise}.}
We display the likelihood evolution under variation of the force parameter in order to visualize the likelihood landscape around this point, even though the models with zero force yield higher evidences. As expected, the force is inferred to be close to zero. 
Evidently the Bayesian method can detect and correctly estimate measurement noise in this example.

As discussed at the end of Subsection \ref{subsec:noise}, an important caveat is that the coordinate dependent mobility conflicts with the assumptions for the noise inference. If the noise is too large, then the mobility at the true position may be poorly estimated by the mobility at the measured position, reducing reliability of the analysis. On the contrary, if the measurement noise is too small to influence the data significantly, the noise will likely not be discovered in smaller data sets. 

\begin{figure}[h]
\begin{center}
\includegraphics[width = 0.45\textwidth]{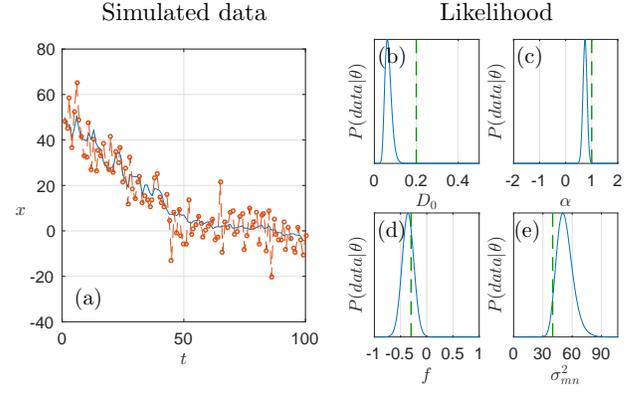}
\caption{(a) The data from Fig.~\ref{sampleplot}(a) is reproduced in solid blue while the orange circles represent data with a large added measurement noise ($\sigma_{\rm mn}^2 = 40$). (b)~-~(e) Likelihoods for variations of single parameters around the maximum likelihood point for the best candidate model ($M_{\text{pull,mn}}$) are shown, displaying inaccuracy caused by the large measurement noise.}
\label{hugenoise}
\end{center}
\end{figure}

To demonstrate possible inaccuracies of the first kind, we infect the data in Fig.~\ref{sampleplot} with a very large measurement noise ($\sigma_{\rm mn}^2 = 40$ in native units) and display the inference results in Fig.~\ref{hugenoise}. 
The large noise hinders the inference of the diffusion constant, while still correctly enables the model comparison (with $Z_{\text{pull,mn}}/Z_{\text{pull,clean}} \sim 10^{18}$) and inference of the noise parameter.

From our experience, the largest inherent inaccuracy we encounter is the statistics itself, i.e., the amount of data points. In addition, if the coordinate does not change by more than an order or magnitude, the exponent $\alpha$ is hard to determine, and therefore all other parameters are inaccurately estimated. This error can be amended, if the system is predicted to have a particular value for $\alpha$. 

\begin{figure}[h]
\begin{center}
\includegraphics[width = 0.45\textwidth]{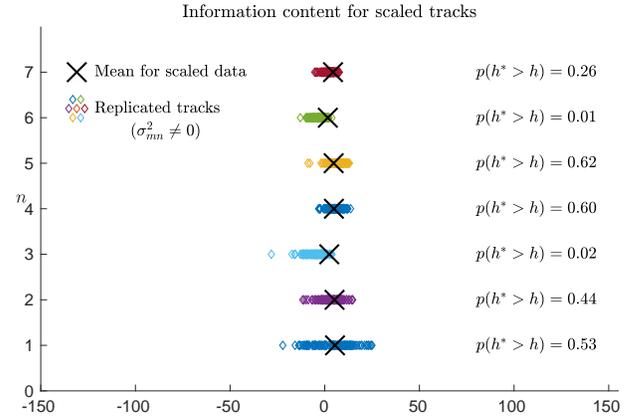}
\caption{Information measures as in Fig.~\ref{inftrue_1} for the data set in Fig.~\ref{hugenoise} assuming the model $M_{\text{pull,mn}}$.}
\label{inf_noise}
\end{center}
\end{figure}

Once again, we compare the inferred information to that of replicated tracks generated from parameters drawn via Eq. \eqref{postdist} for the model with noise. The results are shown in Fig.~\ref{inf_noise}.
The large measurement error effects dominate this data set, and the analysis deems the data set to be typical for the model, providing a check of whether critical aspects have been overlooked.

\begin{figure}[h]
\begin{center}
\includegraphics[width = 0.45\textwidth]{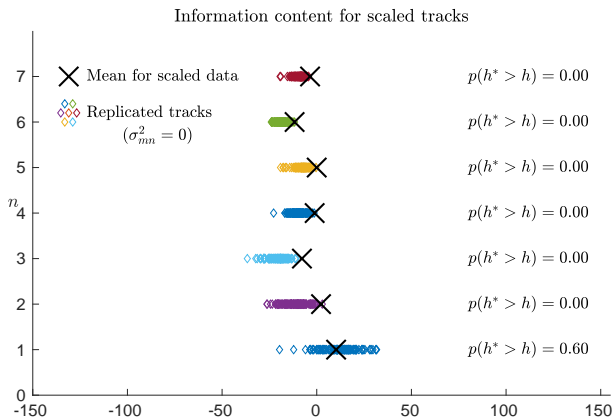}
\caption{Information measures as in Fig.~\ref{inftrue_1} for the data set in Fig.~\ref{hugenoise} assuming the model $M_{\text{pull,clean}}$.}
\label{inf_nonoise}
\end{center}
\end{figure}

In contrast, carrying out the analysis without a measurement error, the information comparison yields the results shown in Fig.~\ref{inf_nonoise}.
In this case, where the measurement error has been overlooked, the replicated data yield different amounts of information than the actual data. Note once again, that this is despite the fact that the Bayesian inference has found the most likely values for the parameters to use for the model. Even though the measurement error in this case would neither produce a mean drift or a relatively substantial contribution to the mean squared displacement at large times, it is revealed by the information comparison and determined by the Bayesian inference.
{\change
Note also that the information content model check for the unscaled data does not reveal that a substantial measurement error has been overlooked. This exemplifies that the undersampling of data can be crucial for the information content model check as discussed further in Appendix \ref{H_scale}.
}

\section{Conclusions}\label{sec:V}
{\change We have demonstrated that} Bayesian inference offers a straightforward method of analysis to systems where forces and/or mobilities are coordinate dependent while also taking into account the stochastic nature of the dynamics. {\change 
This was found to be  contrasted with a simple analysis where deterministic equations of motion were fitted to a trajectory yielding parameter values off the underlying true values.
%These systems are typically not suited for analysis via methods using mean squared displacement or deterministic equations of motion for the average behavior.
 While also being able to compare the relative merit of several models, the Bayesian analysis provides estimates of the parameters for each model via their posterior probability distribution that are consistent with the underlying true values.} 

One limitation of a strict Bayesian approach is that it does not necessarily reveal serious shortcomings of the candidate models, since it only supplies their relative merits. Here, {\change we found that} a combination with a frequentistic approach turns out to be useful. The information content model check {\change combined with undersampling} was shown to be able to detect such shortcomings of models even in the case where the model was the best ``fit". This includes ``obvious" parameters such as an extra force as well as measurement error parameters which are harder to discover by visual inspection. {\change The test will not capture all kinds of shortcomings though. Thus it could be beneficial to supplement it with further model checks. In particular ones that match key properties of the specific system at hand.}

We expect the Bayesian inference coupled with the information content model check approach presented in this article to have many applications within the field of single particle tracking.

\section*{Acknowledgements}

The authors acknowledge support from the Danish Council for Independent Research - Natural Sciences (FNU), grant number 4002-00428B. We thank Tobias Ambj{\"o}rnsson and Christoffer Lagerholm for stimulating discussions.

\appendix

\section{Measurement noise inference}
\label{app_noise}
The conclusions of this section have been drawn in \cite{rasmussen06}. We rederive the results here in order to align the notation.
We assume that the actual coordinate $x^{\text{true}}_i$ is polluted by noise $\eta_i$ to produce the observed coordinates 
\begin{align}
x_i &= x_i^{\text{true}} + \eta_i 
\nonumber  \\
    &= x_{i-1}^{\text{true}} + \Delta x_i^{\text{true}} + \eta_i
\nonumber  \\
    &= x_{i-1} - \eta_{i - 1} + \Delta x_{i}^{\text{true}} + \eta_i.\label{eq:mofx}
\end{align}
where the $\eta_i$ are independent and Gaussianly distributed with zero mean and variance $\sigma_{\rm mn}^2$ and $\Delta x_i^{\text{true}} = x_i^{\text{true}} - x_{i-1}^{\text{true}}$. 

For the displacement $\Delta x_i^{\text{true}}$ we assume that the average $E(\Delta x_i^{\text{true}}) = \Delta\bar{x}$ and variance $\text{Var}(\Delta x_i^{\text{true}}) = \sigma_d^2$ are independent of the coordinates. 
Thus after the first step their conditional covariances are
\begin{align}
\text{Cov}(x_1,x_1|x_0) &= \sigma_d^2 + 2 \sigma_{\rm mn}^2 \nonumber \\
\text{Cov}(x_1,\eta_1|x_0) &= \sigma_{\rm mn}^2 \\
\text{Cov}(\eta_1,\eta_1|x_0) &= \sigma_{\rm mn}^2, \nonumber
\end{align}
and 
\begin{equation}
\text{P}\left(
\begin{bmatrix} \eta_1 \\ x_1
\end{bmatrix} | x_0
\right)
\sim N \left(
\begin{bmatrix} 0 \\ x_0+ \Delta\bar{x}
\end{bmatrix},
\begin{bmatrix} \sigma_{\rm mn}^2 & \sigma_{\rm mn}^2 \\
\sigma_{\rm mn}^2 & \sigma_d^2 + 2\sigma_{\rm mn}^2
\end{bmatrix}
\right).
\label{initp}
\end{equation}
where $N(\vek{\mu},\vek{\Gamma})$ represents a Gaussian distribution with mean $\vek{\mu}$ and covariance matrix $\vek{\Gamma}$. Now assume that we know $\tilde{x}_{i}$ and $\tilde{\sigma}_i^2$ such that
\begin{equation}
\label{pxiandx}
\text{P}\left(
\begin{bmatrix} \eta_i \\ x_i
\end{bmatrix} |\vek{x}_{i-1}
\right)
\sim N \left(
\begin{bmatrix} 0 \\ \tilde{x}_{i}
\end{bmatrix},
\begin{bmatrix} \sigma_{\rm mn}^2 & \sigma_{\rm mn}^2 \\
\sigma_{\rm mn}^2 & \tilde{\sigma}_i^2 
\end{bmatrix}
\right).
\end{equation}
where $\vek{x}_{i-1}$ is a vector containing as elements $x_0,x_1,\dots,x_{i-1}$. To iterate we then need to find $\tilde{x}_{i+1}$ and $\tilde{\sigma}_{i+1}^2$.
We can infer $P(\eta_i|\vek{x}_i)$ via
\begin{equation}\label{peta_bayes}
P(\eta_i|\vek{x}_i) = \frac{P(\eta_i,x_i|\vek{x}_{i-1})}{P(x_i|\vek{x}_ {i-1})}. 
\end{equation}
where
\begin{equation}\label{p_xixi}
P(x_i|\vek{x}_{i-1}) = \frac{1}{\sqrt{2\pi \tilde{\sigma}_i^2}}\text{ exp}\left(-\frac{1}{2\tilde{\sigma}_i^2}(x_i - \tilde{x}_i)^2 \right).
\end{equation}
Combining Eqs. \eqref{pxiandx}, \eqref{peta_bayes} and \eqref{p_xixi}, one obtains the Gaussian probability density function $P(\eta_i|\vek{x}_i)$, from which we read off the conditional mean and variance for $\eta_i$
\begin{align}
E(\eta_i|\vek{x}_i)&=\frac{\sigma_{\rm mn}^2}{\tilde{\sigma}_{i}^2}(x_{i} - \tilde{x}_{i})\\
\text{Var}(\eta_i|\vek{x}_i)&=\sigma_{\rm mn}^2\left(1-\frac{\sigma_{\rm mn}^2}{\tilde{\sigma}_{i}^2} \right)
\end{align}
 to be used with Eq. \eqref{eq:mofx} to obtain
\begin{align}
\tilde{x}_{j+1} &= E(x_{j+1}|\vek{x}_{j}) = x_j +\Delta\bar{x}- \frac{\sigma_{\rm mn}^2}{\tilde{\sigma}_{j}^2}(x_{j} - \tilde{x}_{j})\,, \\ 
\tilde{\sigma}_{j+1}^2 &= \text{Var}(x_{j+1}|\vek{x}_{j}) =  \sigma_d^{2} + \sigma_{\rm mn}^2\left(2-\frac{\sigma_{\rm mn}^2}{\tilde{\sigma}_{j}^2} \right),
\end{align}
with the initial values given in Eq. \eqref{initp}.

\section{Information for scaled tracks}
\label{H_scale}

In this appendix we will demonstrate that diffusive Gaussian models will pass the information content model check if no scaling of steps is performed. To do this we begin by looking at the logarithm of a Gaussian likelihood
\begin{multline}
h(\vek{x^*},\vek{\theta})=\ln P(\vek{x^*}|\vek{\theta},M)-\ln Z(M)\\=-\frac{N}{2}\ln(2\pi)-\frac{1}{2}\ln\det\vek{\Gamma}-\ln Z(M)\\
-\frac{1}{2}(\vek{x^*}-\vek{\mu})^T\vek{\Gamma}^{-1}(\vek{x^*}-\vek{\mu}).\label{eq:hexp}
\end{multline}
Here $N$ is the size of the replicated observations $\vek{x}^*$, $\vek{\Gamma}$ is the model covariance, $\vek{\mu}$ the mean and ${}^T$ denotes transposition. 
We would like to compare the typical $h(\vek{x}^*,\vek{\theta})$ with the corresponding values for the real observations $h(\vek{x},\vek{\theta})$. We will do this by comparing their averages over the random draws of $\vek{x}^*$ and $\vek{\theta}$, i.e., comparing
\begin{equation}
\mathcal{H}^*=\int h(\vek{x^*},\vek{\theta})P(\vek{x}^*|\vek{\theta},M)P(\vek{\theta}|\vek{x},M)d\vek{x}^*d\vek{\theta}\label{eq:Hstar}
\end{equation}
with
\begin{align}
\mathcal{H}&=\int h(\vek{x},\vek{\theta})P(\vek{x}^*|\vek{\theta},M)P(\vek{\theta}|\vek{x},M)d\vek{x}^*d\vek{\theta}\nonumber\\
&=\int h(\vek{x},\vek{\theta})P(\vek{\theta}|\vek{x},M)d\vek{\theta}\label{eq:H}.
\end{align}
Since a difference between $\mathcal{H}^*$ and $\mathcal{H}$ can only arise from the last term in Eq. \eqref{eq:hexp} we will define
\begin{equation}
{\tilde h}(\vek{x^*},\vek{\theta})=-\frac{1}{2}(\vek{x^*}-\vek{\mu})^T\vek{\Gamma}^{-1}(\vek{x^*}-\vek{\mu})
\end{equation}
and compare the corresponding $\tilde{\mathcal{H}}^*$ and ${\tilde{\mathcal{H}}}$ defined similarly to Eqs. \eqref{eq:Hstar} and \eqref{eq:H}.

$\tilde{\mathcal{H}}^*$ is straightforwardly evaluated. If we average ${\tilde h}(\vek{x^*},\vek{\theta})$ over $\vek{x}^*$, indicating this average over the distribution $P(\vek{x}^*|\vek{\theta},M)$ by $\langle\dots\rangle_{\vek{x}^*}$, we get
\begin{multline}
\langle (\vek{x}^*-\vek{\mu})^T\vek{\Gamma}^{-1}(\vek{x}^*-\vek{\mu})\rangle_{\vek{x}^*}=\\
\mathrm{Tr}\left[\vek{\Gamma}^{-1}\langle(\vek{x}^*-\vek{\mu})(\vek{x}^*-\vek{\mu})^T\rangle_{\vek{x}^*}\right]
=\mathrm{Tr}\left[\vek{\Gamma}^{-1}\vek{\Gamma}\right]=N,\label{eq:Hstarcalc}
\end{multline}
since $\vek{\mu}$ and $\vek{\Gamma}$ is the mean and covariance of the components of $\vek{x}^*$. Thus we have that $\tilde{\mathcal{H}}^*=-N/2$.

To evaluate $\tilde{\mathcal{H}}$ we need to specify some of the parameter dependence in the model. Here we will assume that all the elements of the covariance matrix scale in the same way with a single parameter, e.g., a diffusion constant. Thus we can write $\vek{\Gamma}=\lambda \vek{\Gamma}_1$ where $\lambda$ represents the scaling with the parameter and $\vek{\Gamma}_1$ is independent of $\lambda$. Furthermore, we will assume a Jeffreys prior for $\lambda$, i.e., that over a broad range of $\lambda$ we have $P(\vek{\theta}|M)=\lambda^{-1}f(\hat{\vek{\theta}})$, where $f$ is some function of the remaining parameters $\hat{\vek{\theta}}$. The range over which this is true is assumed to be much broader than the peak of the posterior for $\lambda$. With these assumptions we have $\vek{\Gamma}^{-1}=\lambda^{-1}\vek{\Gamma}_1^{-1}$ and we can rewrite $\tilde{\mathcal{H}}$ conveniently as
\begin{align}
\tilde{\mathcal{H}}=&\int {\tilde h}(\vek{x},\vek{\theta})\frac{P(\vek{x}|\vek{\theta},M)P(\vek{\theta}|M)}{Z(M)}d\vek{\theta}\nonumber\\
=&\; Z(M)^{-1}\int \lambda^{-1}f(\hat{\vek{\theta}})\left[(2\pi)^N\lambda^N\det\Gamma_1\right]^{-1/2}\\
&\; \times(-\lambda)\frac{d}{d\lambda}\exp\left[-\frac{1}{2\lambda}(\vek{x}-\vek{\mu})^T\vek{\Gamma}_1^{-1}(\vek{x}-\vek{\mu})\right]d\lambda d\hat{\vek{\theta}}\nonumber
\end{align}
Doing a partial integration, using the assumption of broadness of the prior $\lambda$-range relative to the posterior to discard the boundary terms, we get
\begin{align}
\tilde{\mathcal{H}}
=&\; Z(M)^{-1}\int \left(-\frac{N}{2}\right)\lambda^{-1}f(\hat{\vek{\theta}})\left[(2\pi)^N\lambda^N\det\Gamma_1\right]^{-1/2}\\
&\; \times\exp\left[-\frac{1}{2\lambda}(\vek{x}-\vek{\mu})^T\vek{\Gamma}_1^{-1}(\vek{x}-\vek{\mu})\right]d\lambda d\hat{\vek{\theta}}\nonumber
\end{align}
If we take the factor $-N/2$ out of the integral, then the remaining integral becomes exactly the evidence $Z(M)$. Thus we have that $\tilde{\mathcal{H}}=-N/2$ just like for $\tilde{\mathcal{H}}^*$.

We have thus found that on average the replicated observations $\vek{x}^*$ will have the same information content as the real observations for this class of Gaussian models. We therefore extend the test by scaling the time steps, i.e., omitting data points. To see why this can make a diffusive Gaussian model fail, we can look at deterministic motion with constant velocity, i.e., steps given by $\Delta x_i=v\Delta t$, where $v$ is the velocity and $\Delta t$ the duration of the steps. If we apply the above test with Brownian motion without drift as the model then we would get
\begin{equation}
\tilde{h}(\vek{x},\vek{\theta})=-\frac{1}{2}\sum_{i=1}^N\frac{\Delta x_i^2}{2 D\Delta t}=-\frac{1}{2}N\frac{v^2\Delta t}{2 D}
\end{equation}
where $D$ is the diffusion constant of the Brownian diffusion model. As the derivation above shows, then the Bayesian analysis will provide a posterior for $D$ that is peaked around $D=v^2\Delta t/2$, such that $\tilde{h}(\vek{x},\vek{\theta})$ has typical values around $-N/2$. However, if one now only keeps every $n$'th data point, then one will have a step duration of $\Delta t'=n\Delta t$ with a total of $N'=N/n$ steps. This will give values of $\tilde{h}(\vek{x}_n,\vek{\theta})$ for the remaining data points $\vek{x}_n$ that are unchanged in this case, since $N'\Delta t'=N \Delta t$. But the values of $\tilde{h}(\vek{x}^*_n,\vek{\theta})$ will on average be reduced by $n$ since  $\tilde{\mathcal{H}}^*_n=-N'/2$ for the reduced replicated data sets $\vek{x}^*_n$ (this follows from a calculation similar to Eq. (\ref{eq:Hstarcalc})). Thus the predicted scaling of the step deviation does not match the scaling inherent in the observations, and therefore Brownian motion without drift will fail the model check when the scaling of the data is included in the check.

%\bibliography{../../../synced/mycites/mycites}

\bibliography{refs}

\end{document}